\documentclass[nofootinbib,floatfix,twocolumn]{revtex4}

\usepackage{empheq} 
\usepackage{mathrsfs}
\usepackage{verbatim}
\usepackage{amssymb}
\usepackage{graphicx}
\usepackage{subfigure}
\usepackage{color}
\usepackage{rotating}

\usepackage[mathscr]{euscript}
\usepackage{amsmath}

\usepackage{mathrsfs}
\usepackage{verbatim}
\usepackage{amssymb}
\usepackage{graphicx}
\usepackage{subfigure}
\usepackage{color}
\usepackage{rotating}

\begin{document}

\title{Energy Level Diagrams for Black Hole Orbits}

\author{Janna Levin}
\affiliation{Department of Physics and Astronomy, Barnard
College of Columbia University, 3009 Broadway, New York, NY 10027 }
\affiliation{Institute for Strings, Cosmology and Astroparticle
  Physics (ISCAP), Columbia University, New York, NY 10027}

\widetext

\begin{abstract}

\centerline{janna@astro.columbia.edu}

\bigskip

A spinning black hole with a much smaller black hole companion forms a
fundamental gravitational system, like a colossal classical analog 
to an atom. In an  
appealing if imperfect analogy to atomic physics, this gravitational
atom can be understood through a discrete spectrum of periodic
orbits. Exploiting a correspondence between the set of periodic orbits
and the set of rational numbers, we are able to construct periodic
tables of orbits and energy level
diagrams of the accessible states around black holes. 
We also present a closed form expression for the rational $q$, thereby
quantifying zoom-whirl behavior
in terms of spin, energy, and angular momentum.
The black hole
atom is not just a theoretical construct, but corresponds
to extant astrophysical systems 
detectable
by future gravitational wave observatories.
\end{abstract}

\maketitle

Bare black holes are as perfect as fundamental particles. As
Chandrasekhar said, ``The black holes of nature are the most perfect
macroscopic objects there are in the universe \cite{chandrasekhar1983}''.
A black hole with a given mass and spin is indistinguishable from every other
black hole with the same mass and spin. 
Likewise, a supermassive black hole with a much smaller black hole companion 
forms a kind of macroscopic, classical atom, reminiscent of the
hydrogren atom. In analogy
with atomic physics, the orbits around a given black hole can be
completely described by a periodic table \cite{levin2008} 
-- a table of periodic orbits ordered
in ascending energy from the stable circular orbits (the ground-like
state) up to the last bound orbits (the energy of ionization).
Further, the energy levels of the periodic orbits around a black hole are,
formally speaking, discrete. In the spirit of this atomic analogy \cite{Drasco:2007gn},
we exploit energy level 
diagrams as another valuable representation of the dynamics
around a black hole nucleus.

\begin{figure}
\begin{minipage}{110mm}
\hspace{-70pt}
 \includegraphics[width=0.225\textwidth]{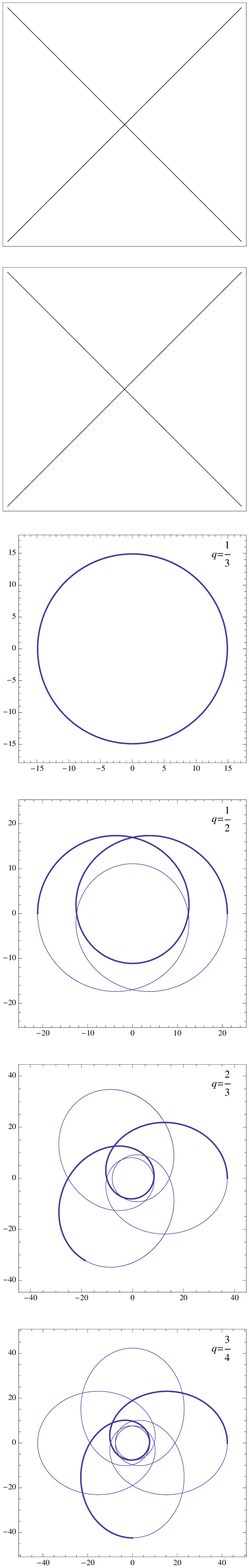}
  \centering
  \includegraphics[width=0.225\textwidth]{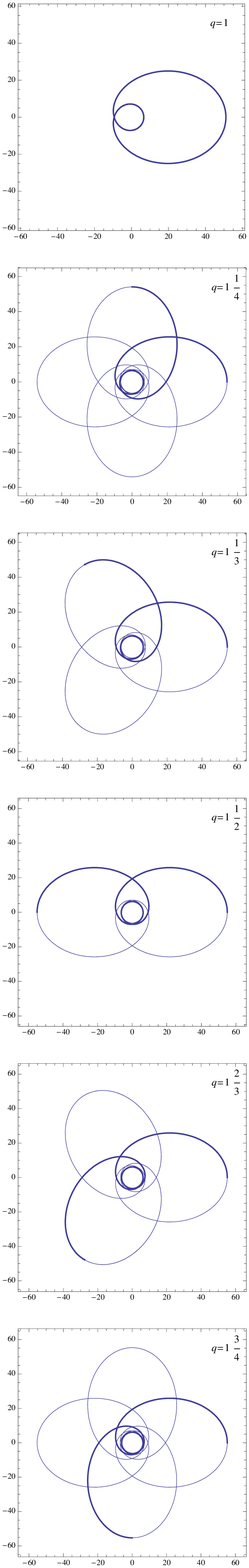} 
  \includegraphics[width=0.225\textwidth]{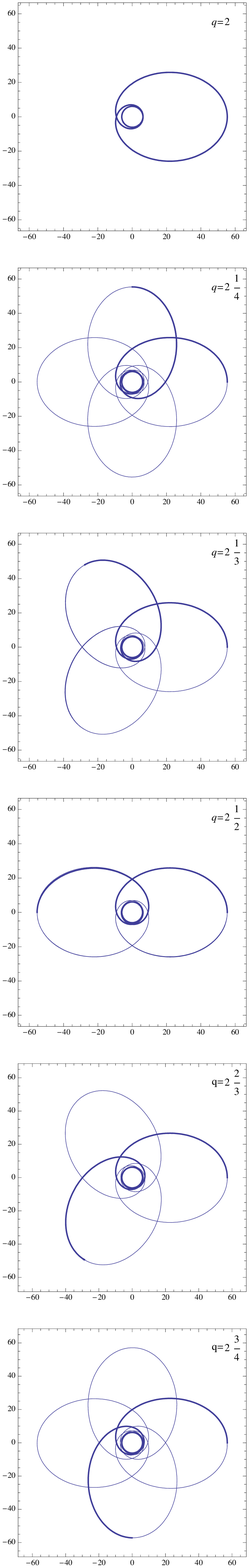}
\hfill
\end{minipage}
  \caption{
A periodic table for retrograde equatorial orbits around a
black hole
with spin parameter $a=-0.9$. 
The angular momentum is $L=4.56056$.
All entries with $z\le 4$ and $0\le w\le
 2$ are included. Energy increases from top to bottom and then left to
 right, as does the rational $q=w+v/z$. The columns are $w$-bands 
 $0,1,2$ from left to right. The minimum is $q\ne 0$, as discussed in \S
 \ref{sec:energy} and \S \ref{closedform}. Although we have drawn 3 columns,
there are in principle an infinite
 number as $w\rightarrow \infty$. }
  \label{retro}
\end{figure}

The energy level diagrams are based on an infinite ordered
set of periodic orbits around a Kerr black hole \cite{levin2008}.
Fig.\ \ref{retro} shows a periodic table of retrograde orbits.  
At first glance, the set of idealized, closed orbits might seem improbably
special.
% since it is a set of measure zero in the space of all possible
% orbits. 
Yet the periodic set is
fundamental to an
understanding of the entire
dynamics. As we now elaborate,
{\it all} orbits are near a periodic orbit, and therefore this
seemingly special set is able to 
encode the entire orbital dynamics.

\section{Periodic Orbits and the Rational Numbers}

Our atomic analogy will stem from a mapping between black hole
 orbits and rational numbers.
 For the sake of argument,
we consider orbits confined to the equatorial plane.
Any equatorial orbit around a black hole has two fundamental
frequencies\footnote{Non-equatorial orbits have three fundamental 
  frequencies and therefore are designated by two rationals.}
-- the libration in the radial coordinate, $\omega_r$,
and the rotation
in the angular coordinate, $\omega_\varphi$.
To be clear, by $\omega_r$ we mean the frequency of radial
oscillations, $\omega_r=2\pi/T_r$, where $T_r$ is the time it takes to
move from one apastron to another. By $\omega_\varphi$, we mean the
average angular frequency over a radial oscillation,
$\omega_\varphi={T_r^{-1}}\int_{0}^{T_r}(d\varphi/dt) dt$.
A generic
orbit will have frequencies that are irrationally related
and will consequently never close. Instead, it precesses
around the central black hole nucleus, never returning exactly to its
starting point. In contrast, a special set of orbits -- the periodic
orbits -- will close. 
For this to be the case, the two fundamental
frequencies must be commesurate,
that is,
rationally related. We define the rational number $q$ associated with
each periodic orbit through
\begin{equation}
\frac{\omega_\varphi}{\omega_r}=1+q \quad\quad .
\end{equation}
 Since the set of irrationals is
a larger infinity than the set of rationals, the rationals are a set
of measure zero in the space of all possible orbits.

However, the set of
periodic orbits are dense on the number
line.
Any irrational can be arbitrarily well
approximated by a nearby rational. A good way to find the rational
approximant is to use continued fractional expansions. Let $\sigma$ be
some irrational number and let $1+\sigma$
be the irrational ratio of frequencies for some generic orbit.
Such a trajectory precesses around
the central black hole without ever closing. The
irrational can be represented through continued fractions as
\begin{equation}
\sigma = w +\frac{1}{a_1+\frac{1}{a_2+\frac{1}{a_3+...}}} 
\end{equation}
where $w$ is the integer part of $\sigma$, $a_1$ is the integer part
of the reciprocal of $\sigma -w$, $a_2$ is the integer part of the
reciprocal of the remainder of $a_1$ etc.\ through an infinite list
of integers $a_n$. A given rational approximant is obtained
by truncating the continued fraction at some finite $n$.
For instance, if we truncate the continued fractional expansion at
$n=3$, then 
\begin{equation}
\sigma \approx q=w+\frac{v}{z}=w+\frac{a_2a_3
  +1}{a_1a_2a_3+a_1+a_3} \quad\quad . 
\label{approxdef}
\end{equation}
In this way, any irrational can be
arbitrarily well approximated by rationals; significantly then,
any aperiodic orbit (with an irrational frequency ratio) can be
arbitrarily well approximated by some 
periodic orbit (with a rational frequency ratio).

\begin{figure}
  \centering
\includegraphics[width=40mm]{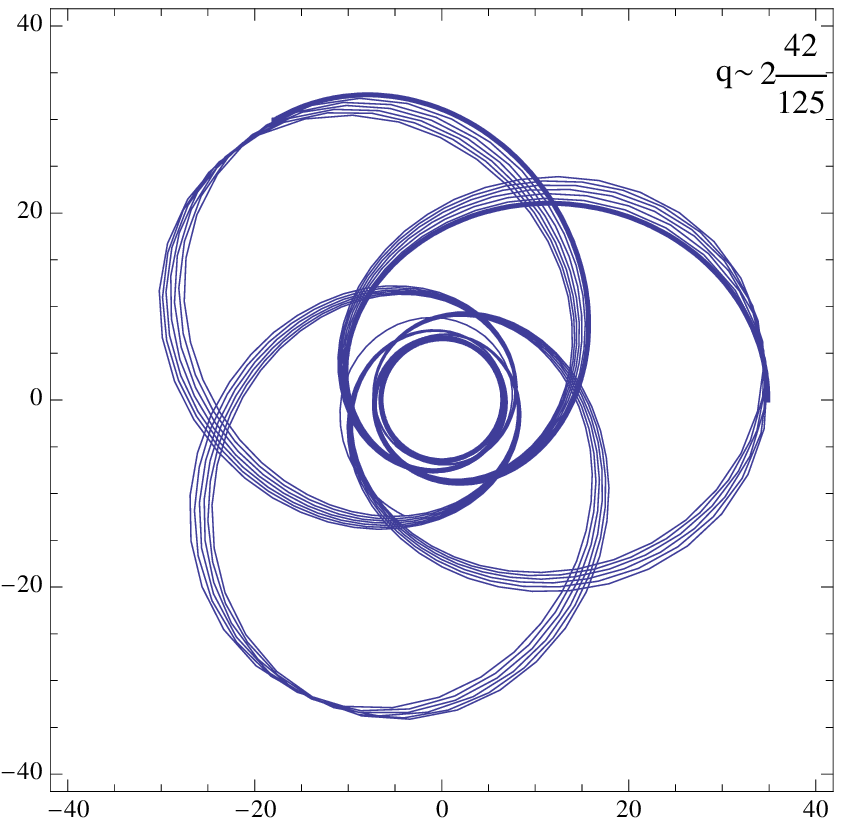}
\includegraphics[width=40mm]{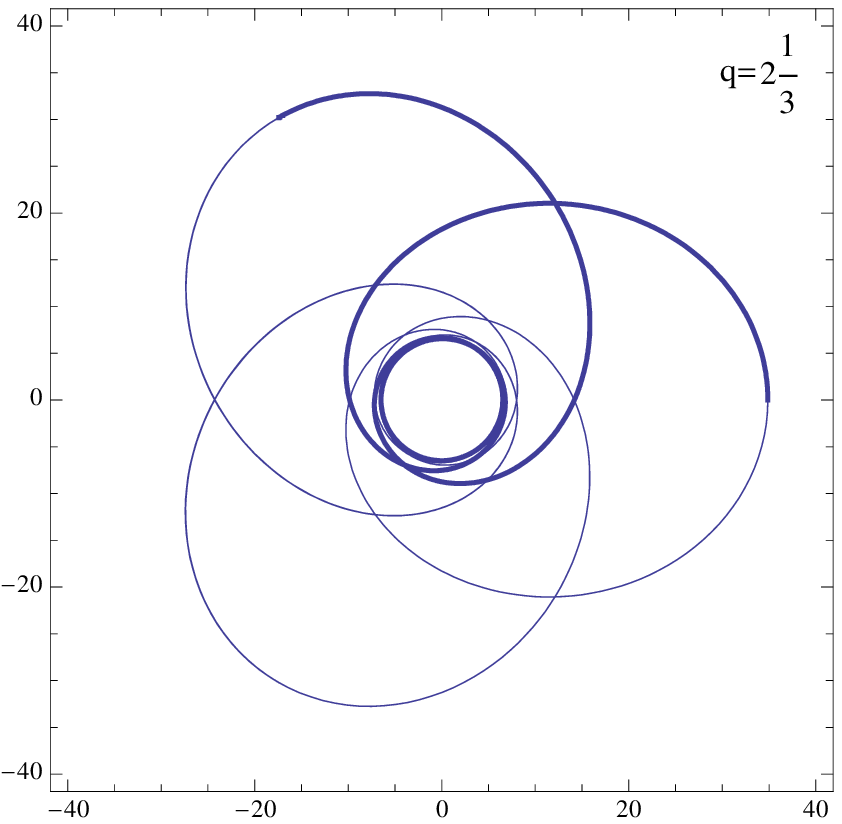}
\hfill
\caption{Left: A generic retrograde orbit 
around
  a central 
  black hole with spin 
  parameter $a=-0.995$. 
The angular momentum $L$ is halfway between the ibco (innermost bound
circular orbit) and the isco (innermost stable circular orbit).
This orbit has an energy very near the $3$-leaf
  orbit on the right.
\label{precess}}  \end{figure}

To illustrate, the generic aperiodic orbit on the left of Fig.\
\ref{precess}, if allowed to run indefinitely, would fill an annulus in
the plane, never closing. This is a generic example of zoom-whirl
behavior. (For other discussions of zoom-whirl behavior see \cite{2002PhRvD..66d4002G}.) 
Yet this same orbit has a $\sigma \approx q=2+42/125$, and so is very close
to a periodic orbit that whirls 2 times, has 125 leaves but skips to
the 42nd next leaf in the pattern each successive time it hits apastron. Also
note that the orbit is still close to, but not as well approximated
by, the much lower-leaf orbit characterized by the rational
$q=2+1/3$, 
drawn
on the right of Fig.\ \ref{precess}. 
Indeed, the
generic orbit can be thought of as a precession of
this fundamental three-leaf pattern by the amount $2\pi(42/125 -
1/3)=2\pi/375$ -- so less than one degree -- each
radial cycle.  In fact, in general,
all high-leaf orbits can be understood as precessions
of lower-leaf orbits \cite{levin2008}.

For emphasis, every orbit that is measured observationally or
generated numerically is necessarily truncated by a rational
approximant and is therefore, strictly speaking,
periodic, albeit possibly with an extremely long
period. Consequently the orbital dynamics around the central 
black hole is entirely defined by the periodic skeleton. Furthermore,
high-leaf periodics can be understood as precessions of lower-leaf periodics.

Although we restrict to
two-dimensional equatorial
motion here, generic three-dimensional non-equatorial orbits can also be identified with
rational numbers. In principle, two rationals are needed to
characterize a completely closed non-equatorial orbit, the $q$ we are using as well
as a $q_\theta=\omega_\theta/\omega_r$.
However, it should be mentioned that we do not need to require fully
closed motion,
as shown in
Refs.\ \cite{levin2008:2,grossman2008} in the context of comparable
mass binaries. The useful insight is the simple observation that
every orbit lies in its own orbital plane, even if it does not
lie in the equatorial plane. Periodic tables of {\it orbital plane}
motion were shown to taxonomize orbits with one rational, $q_\Phi$
where $\Phi$ is the angle swept out in the orbital plane, 
just as the periodic tables of equatorial orbits are
taxonomized by the $q$ above. The entire orbital plane then precesses
to trace out fully three-dimensional motion, although not necessarily
fully periodic motion.
For the rest of this paper, we will restrict to equatorial motion for transparency.

\section{$q$ and its Topological Significance}

We can relate the triplet of integers $(z,w,v)$ in the definition of
the rational 
\begin{equation}
q=w+v/z
\label{rationaldef}
\end{equation}
to
geometric and topological features of the orbit. Consider Fig.\
\ref{retro} again. The periodic orbits are assembled in a table of
increasing energy -- as is the
chemical periodic table. In our tables, energy increases
from top to bottom and left to right. Note that the $q$ of the periodic table
increases monotonically with energy. 

Each $q$ also has a
geometric interpretation. Consider its definition
\begin{equation}
q=w+\frac{v}{z}\equiv \frac{\omega_\varphi}{\omega_r}-1 =\frac{\Delta
  \varphi}{2\pi}-1
\label{qequiv}
\end{equation}
where $\Delta \varphi=\int^{T_r}(d\varphi/dt)dt $ is the equatorial
angle accumulated in one 
radial cycle from apastron to apastron. By this definition, we see
that $q$ is the amount an
orbit precesses beyond the closed ellilpse. Therefore Kepler's
elliptical orbits have $q=0$ since there is no precession.
Keplerian orbits are all periodic. They are also
relativistically excluded. Relativistic orbits around black holes all
overshoot their starting point, which is to say they all precess, a
discovery that originates with Mercury's famed precession, and all
relativistic orbits have a $q>0$. 

Note that all of the orbits in the second column of the periodic
table of Fig.\ \ref{retro} whirl an 
additional $2\pi$ 
in a nearly circular loop around the nucleus before moving out to the
next apastron. The entries in the third column whirl twice.
The integer 
part of $q$ from equation (\ref{qequiv}), denoted by $w$, is
the number of 
whirls. The periodic tables are ordered in terms of $w$-bands,
which form energy bands. In fact, unlike the chemical periodic table,
the black hole periodic table in Fig.\ \ref{retro} has infinitely many columns, one for
each possible integer number of whirls.

Within each energy band, each $w$-band, the
entries are $z$-leaf orbits that whirl $w$ times between each apastron.
Closed orbits can skip leaves each radial period.
For example, the 
third and fifth entries of column 2 in the periodic tables are both
3-leaf clovers but they are not identical. The 
third
entry in column 2 moves to the next apastron in the $3$-leaf pattern
as indicated in bold.
However, the fifth entry in column 2 moves to the second
apastron in the $3$-leaf pattern as indicated in bold.
According to equation (\ref{qequiv}), the $q$
of the former closed orbit is 
$q=w+v/z=1+1/3$ while the $q$ of the latter is $q=w+v/z=1+2/3$. So $v$
labels the order in which the leaves are hit.

The rational number identifying a given periodic orbit therefore encodes
significant topological and geometric information. It identifies the
number of leaves, the number of whirls, and the order in which the
leaves are traced.

\section{Energy Level Diagrams}
\label{sec:energy}

\begin{figure}
  \centering
  \includegraphics[width=0.3\textwidth,angle=-90]{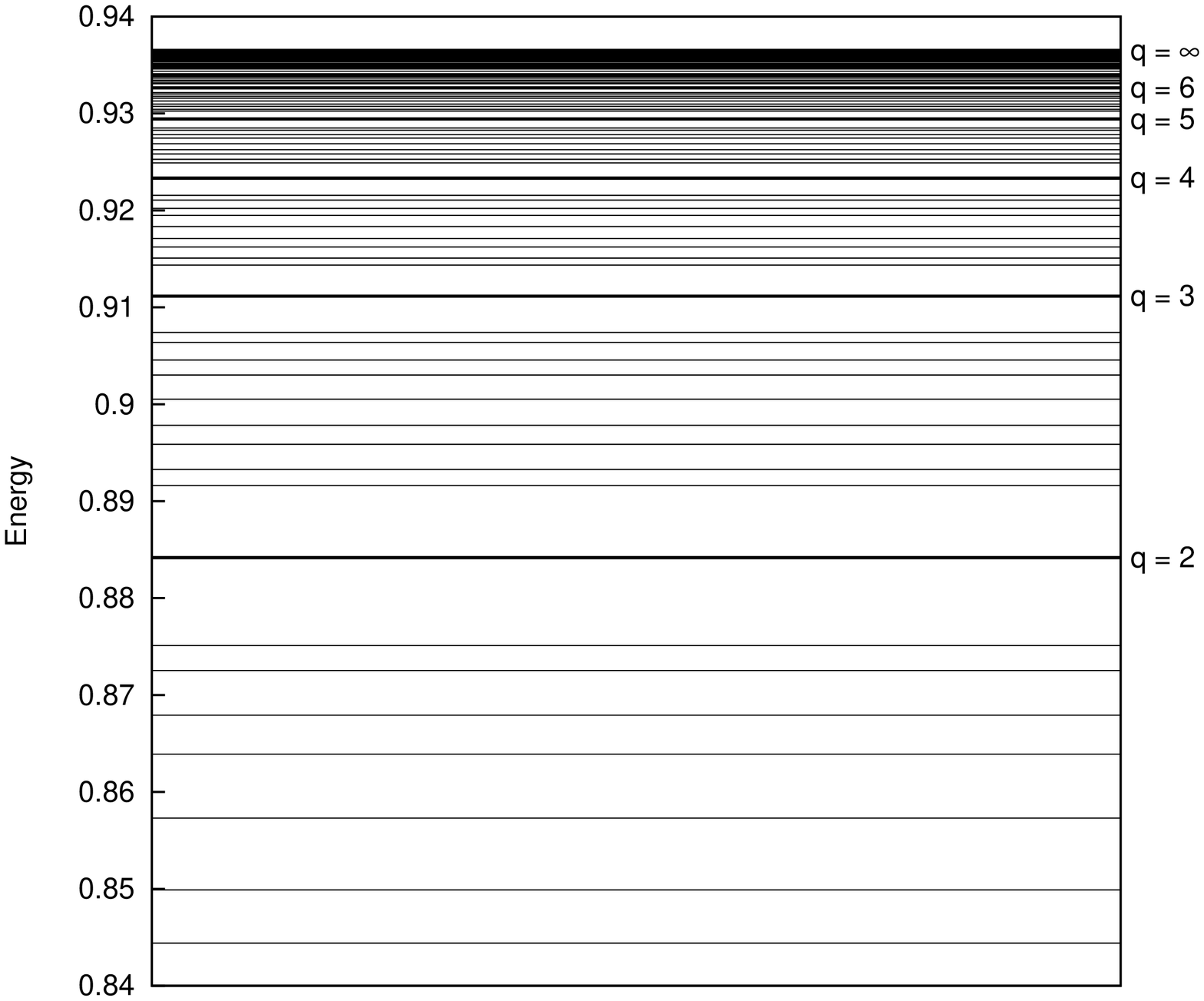}
  \includegraphics[width=0.3\textwidth,angle=-90]{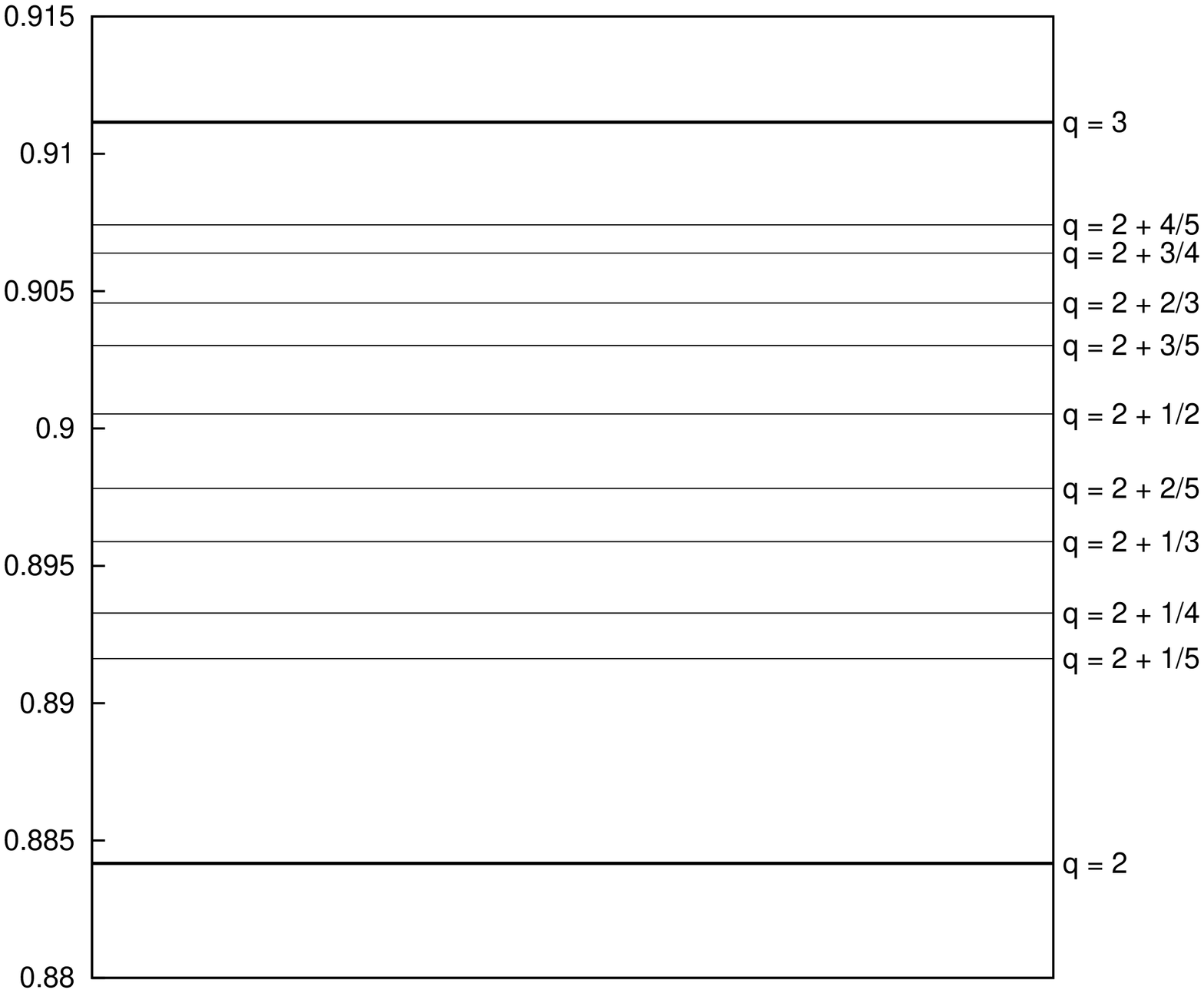}
\hfill
\caption{Energy Level diagrams for equatorial orbits with $L=2$ around
  a central 
  black hole with spin 
  parameter $a=0.995$. All $z\le 5$ for all $1\le w\le 12$ are
  included although the $q=1\frac{1}{5}$
  and $q=1\frac{1}{4}$ entries are missing from the $w=1$ band since
  they are not accessible. The figure on the bottom shows a detail of
  the top figure,
  indicating the rationals within the $w=2$ band. 
\label{energy}}  \end{figure}

Since the rationals are discrete, it follows that
the energies of the periodic orbits are likewise discrete.
Furthermore, as
discovered in Ref.\ \cite{levin2008}, $q$ is monotonic with energy
(for a given angular momentum $L$ around a given black hole). This
latter observation allows us to represent the information in a
periodic table in a related energy level diagram as shown in Fig.\
\ref{energy}.

Energy level diagrams provide a valuable complement to the periodic
tables, effectively and concisely summarizing information. For
instance, Fig.\ \ref{energy} incorporates all
$z\le 5$ orbits for $1\le w\le 12$, while the corresponding periodic
table of Fig.\ \ref{retro} only includes $z\le 4$ for $0\le w \le 2$.

The energy level diagram illustrates the stacking of energy levels for
all orbits 
into bulk $w$-bands. The
fractional part of $q$, represented by $v/z$, has an intriguing
distribution within a given band that roughly repeats
in a scaled manner
throughout all $w$-bands.

If we were to draw every one of the infinite $q$'s allowed, they would,
despite
being discrete,
fill a solid wall of color for each $w$-band. Still,
even though there are always an infinite number of entries in any
periodic table, high-$z$ orbits can be approximated as precessions
arounds low-$z$ clovers. In practice, we can restrict to a finite set
of discrete rationals.

Furthermore, not all rationals are accessible for a given angular
momentum, $L$,
around a given black hole. 
The lower limit is set by the
$q$ of the ground 
state -- equivalently, the $q$ of the stable circular orbit -- and the
upper limit is set by the $q$ of the highest-energy bound orbit.
Both upper and lower limits
depend
on $L$ and the spin of the central black hole.

Although initially counter-intuitive, the lower limit is not  
simply $0$. As in Ref.\ \cite{levin2008}, we take the $q$ for the
ground state, that is, for a
given stable circular orbit, to be $q_c\rightarrow
\omega_\varphi/\omega_r -1 $ in the limit of eccentricity going to
zero. This minimum $q$ is not zero.
For the energy level
diagram of Fig.\ \ref{energy}, the bound on $q$ is roughly 
$1+1/3<q<\infty$.
Importantly, this means that some low $q$ orbits are not accessible in
the strong-field regime. Namely, Mercury-type orbits, which as
precessions of Keplerian ellipses would correspond to rationals
$0<q\ll 1$,
are forbidden in 
the strong-field regime and {\it all} allowed states will look like
precessions of multi-leaf orbits.
We will also write down an explicit analytic expression
for $q_c$ in \S \ref{closedform}.

As the angular momentum increases both the lower and
upper bounds will eventually decrease toward zero, so that, as we know
from solar system dynamics,
all orbits in the
Newtonian approximation are tight precessions of the ellipse.

\section{Closed form expression for $q(a,E,L)$}
\label{closedform}

We can always numerically integrate the Kerr equations to find
$q$. However, it is useful to have a closed form expression purely in
  terms of $(a,E,L)$. Such a closed form should be advantageous when
considering the evolution of $q$ under the dissipative effects of
gravitational radiation \cite{Healy:2009zm}.
Starting with the definition
\begin{equation}
q=\frac{2}{2\pi}\int_{r_p}^{r_a}\frac{\dot \varphi}{\dot r} dr - 1 \quad,
\label{qstart}
\end{equation}
we can
perform this integral to generate an expression in terms of elliptical integrals.
We will take
the further step of finding the limits of integration in terms of
$(a,E,L)$ so that our final expression will automatically deliver $q$
given $(a,E,L)$ as the only input.

We rewrite \cite{levin2008}
the geodesic equations for Kerr equatorial orbits \cite{carter1968}
as 
\begin{eqnarray}
r^2 \dot r &=& \pm \sqrt{R} \nonumber \\
r^2 \dot \varphi &=&-\frac{1}{\Delta}\frac{\partial R}{\partial L}
\label{dimcarter}
\end{eqnarray}
where an overdot denotes
differentiation
with respect to the particle's proper time $\tau$ and
\begin{eqnarray}
R(r) &=& P^2-\Delta \left\{  r^2+(L-aE)^2 \right\} \nonumber \\
P(r) &=& E(r^2+a^2)-aL \nonumber \\
\Delta &\equiv & r^2-2r+a^2 \quad\quad ,
\label{dimpots}
\end{eqnarray}
where the mass of the black hole is set to $M=1$.

We then have
\begin{equation}
q=\frac{1}{\pi}\int_{r_p}^{r_a}\left (\frac{-\left
(aE-L\right )+\frac{a}{\Delta}P}{\sqrt{R}} \right )dr -1 \quad,
\end{equation}
where we take the plus sign in $\dot r$ indicating we are integrating
from periastron, $r_p$, out to apastron, $r_a$.

To solve this, we write both $R$ and
$\Delta$ in terms of their
roots. 
$R$ is a fourth-order equation in $r$. 
In the equatorial plane, one of
the roots is always $0$ and $R$ can be written
\begin{equation}
R(r)=(E^2-1)r(r-r_0)(r-r_p)(r-r_a)
\end{equation}
while we can write $\Delta$ in terms of the inner and outer
horizons $r_{\pm}=1\pm\sqrt{1-a^2}$:
\begin{equation}
\Delta=(r-r_+)(r-r_-) \quad .
\end{equation}

We want to solve
\begin{align}
q  =-\frac{1}{\pi}(aE-L)\int_{r_p}^{r_a}\frac{dr}{\sqrt{R}}
&+\frac{1}{\pi}a^2(aE-L)\int_{r_p}^{r_a}\frac{dr}{\Delta\sqrt{R}}\nonumber \\
&+\frac{1}{\pi}aE\int_{r_p}^{r_a}\frac{r^2dr }{\Delta\sqrt{R}} -1 \ ,
\end{align}
which we rewrite as
\begin{align}
q&=\left [-\frac{1}{\pi}(aE-L){\cal I}_1 +\frac{1}{\pi}a^2(aE-L){\cal I}_2
+\frac{1}{\pi}aE{\cal I}_3\right ]_{r_p}^{r_a} - 1\nonumber \\
{\cal I}_1 &=\int \frac{dr}{\sqrt{R}}\nonumber \\
{\cal I}_2 &=\int \frac{dr}{\Delta\sqrt{R}}\nonumber \\
{\cal I}_3&=\int \frac{r^2dr }{\Delta\sqrt{R}} \quad .
\end{align}
Defining
\begin{align}
\alpha(r) &=\frac{r (r_p-r_0)}{ r_p(r-r_0)}\nonumber \\
\beta(r)  &=\frac{r (r_p-r_a)}{ r_p(r-r_a)} \nonumber\\
\gamma(r)&=\frac{(r-r_0)(r_a-r_p)}{ (r_a-r_0)(r-r_p)}
\end{align}
the solutions are
\begin{widetext}
\begin{align}
{\cal I}_1(r)&=\frac{2  }{\sqrt{(E^2-1)r_p(r_0-r_a)}}
\text{EllipticF}\left[\text{ArcSin}
\left[\frac{1}{\sqrt{\gamma(r)}}\right],
\beta(r_0)\right]
\nonumber\\
{\cal I}_1(r_a)&= \frac{2}{\sqrt{(E^2-1)r_p(r_0-r_a)}}{\text{EllipticK}}
\left [\beta(r_0) \right ]
\end{align}
with ${\cal I}_1(r_p)=0$;
The second integral is
\begin{align}
{\cal I}_2=
\frac{-2 }{{a^2}\sqrt{(E^2-1)r_p (r_a-r_0)} }
\left(\right.&\left.
\text{EllipticF}\left[\text{ArcSin}\left[\alpha(r)^{-1/2}\right],\alpha(r_a)\right]
\right. \nonumber \\ & \left.
+\frac{r_0r_+}{
(r_0-r_-) 
(r_--r_+)}
 \text{EllipticPi}\left[\alpha(r_-),
   \text{ArcSin}\left[\alpha(r)^{-1/2}\right],\alpha(r_a)\right]
\right. \nonumber \\ &\left.
+
\frac{r_0r_-}{(r_0-r_+)
(r_+-r_-)}
 \text{EllipticPi}\left[\alpha(r_+),
   \text{ArcSin}\left[\alpha(r)^{-1/2}\right],\alpha(r_a)\right]
\right)
\nonumber
\end{align}
This gives non-zero contributions at both limits of integration;
Finally,
we have
\begin{align}
{\cal I}_3=&
\frac{2 (r_p-r_0)}
{ \sqrt{(E^2-1)r_p(r_0-r_a)}}
\nonumber \\
&\left(\frac{r_0^2}{(r_0-r_-)
(r_p-r_0) (r_0-r_+)}
    \text{EllipticF}\left[\text{ArcSin}\left[\frac{1}{\sqrt{\gamma(r)}}\right],\beta(r_0)\right]\right.\nonumber \\
&\left.+\frac{r_-^2 }{(r_0-r_-) (r_--r_p)
  (r_--r_+)}
\text{EllipticPi}\left[\gamma(r_-),\text{ArcSin}\left[\frac{1}{\sqrt{\gamma(r)}}
\right],\beta(r_0)
\right]+\right.\nonumber \\
&\left.\frac{r_+^2}{(r_0-r_+) (r_+-r_p)
  (r_+-r_-)}
\text{EllipticPi}\left[\gamma(r_+),\text{ArcSin}\left[\frac{1}{\sqrt{\gamma(r)}}
\right],\beta(r_0)\right]
\right)
\end{align}
At $r=r_a$, $\gamma(r_a)=1$ and an $\text{EllipticF}$ becomes an
$\text{EllipticK}$, while
at $r=r_p$, $\gamma(r_p)=\infty$ and $\text{EllipticF}$ and
$\text{EllipticPi}$ become zero. So evaluating this at the upper and
lower limits gives
\begin{align}
{\cal I}_3(r_a)=&
\frac{2 (r_p-r_0)}
{ \sqrt{(E^2-1)r_p(r_0-r_a)}}
\nonumber \\
&
\left(\frac{r_0^2}{(r_0-r_-)
(r_p-r_0) (r_0-r_+)}
    \text{EllipticK}\left[\beta(r_0)\right]\right.\nonumber \\
&\left.+\frac{r_-^2 }{(r_0-r_-) (r_--r_p)
  (r_--r_+)}
\text{EllipticPi}\left[\gamma(r_-),\beta(r_0)
\right]+\right.\nonumber \\
&\left.\frac{r_+^2}{(r_0-r_+) (r_+-r_p)
  (r_+-r_-)}
\text{EllipticPi}\left[\gamma(r_+),\beta(r_0)\right]
\right)
\quad .
\end{align}

We use these to evaluate
\begin{equation}
q =-\frac{1}{\pi}(aE-L){\cal I}_1(r_a) +\frac{1}{\pi}a^2(aE-L)\left [{\cal I}_2(r_a)-{\cal
    I}_2(r_p)\right ] +\frac{1}{\pi}aE{\cal I}_3(r_a) - 1
\label{qfinal}
\end{equation}
We are not quite done since this expresses $q(a,E,L,r_0,r_p,r_a)$.
To have $q$ as a function of $(a,E,L)$ only, we have to find $r_0,r_p,r_a$ as
functions of $(a,E,L)$. 
To do so we expand the polynomial
$R(r)=(E^2-1)r(r-r_0)(r-r_p)(r-r_a)$ and equate to the definition of $R(r)$
in Eq.\ (\ref{dimpots}), matching up coefficients in powers of $r$ and
finding a system of equations for $r_0,r_p,r_a$. Since $r=0$ is always
a root, this is equivalent to a 3rd order equation in $r$
 and cubic
equations have a generic solution. We want the roots of 
\begin{align}
R(r)&=0\nonumber \\
Ar^3+Br^2+Cr+D &=0\nonumber \\
A &= (E^2-1)\nonumber \\
B &= 2\nonumber \\
C &=A a^2-L^2\nonumber \\
D &=2(L-aE)^2
\label{coeff}
\end{align}
The nonzero roots in ascending order are 
\begin{align}
r_0&=-\frac{B}{3A}
-\frac{2^{1/3}(-B^2+3AC)}{3A}(F+\sqrt{G})^{-1/3}
+\frac{1}{3A2^{1/3}}(F+\sqrt{G})^{1/3}
\nonumber \\
r_p&=-\frac{B}{3A}
+\frac{(1-i\sqrt{3})(-B^2+3AC)}{3A2^{2/3}}(F+\sqrt{G})^{-1/3}
-\frac{1+i\sqrt{3}}{6A2^{1/3}}(F+\sqrt{G})^{1/3}
\nonumber \\
r_a&=-\frac{B}{3A}
+\frac{(1+i\sqrt{3})(-B^2+3AC)}{3A2^{2/3}}(F+\sqrt{G})^{-1/3}
-\frac{1-i\sqrt{3}}{6A2^{1/3}}(F+\sqrt{G})^{1/3}
\label{roots}
\end{align}
where
\begin{align}
F&=  -2B^3 + 9ABC-27A^2D\nonumber \\
G&={F^2-4(B^2-3AC)^3} \quad\quad.
\end{align}
Using Eqs.\ (\ref{roots}) in Eq.\ (\ref{qfinal}), gives the rational $q$ as
a function of $(a,E,L)$, which is what we were after. 

To be clear, the function $q(a,E,L)$ of Eq.\ (\ref{qfinal}) can in
principle be any
positive real number. The periodic orbits correspond to rational $q$. All other
orbits can be arbitrarily well approximated by a rational $q$ as
discussed previously.
As a check, inputting the $(a,E,L)$ of the retrograde orbit on the right
of Fig.\ \ref{precess} returns a $q=2.3333...=2\frac{1}{3}$, as it should.

\bigskip

\end{widetext}

While Eq.\ (\ref{qfinal}) gives the $q$ for a generic equatorial orbit,
it is interesting to write out the
expression for the lower limit of stable 
circular orbits explicitly. Circular orbits have a radius
$r_c=r_a=r_p$ and this leads to a great
simplification in the expression since then
$\beta(r_0)=\gamma(r_+)=\gamma(r_-)=0$ and 
$\text{EllipticK}[0]=\text{EllipticPi}[0,0]=\pi/2$.
\begin{comment}
Also notice 
\begin{equation}
\frac{r_+^2 - r_-^2}{(r_+ - r_-)}= 2\quad .
\end{equation}
\end{comment}
Using these relations gives
\begin{align}
q_c=&\frac{1}{\sqrt{(E^2-1)r_c(r_0-r_c)}}
\left [-(aE-L)
+aE\frac{r_c^2}{\Delta(r_c)}
 \right ]\nonumber 
\end{align}
The root $r_0$ corresponding to a given $r_c$ can be found
from the cubic expressions above evaluated at the energy and
angular momenta of circular orbits, originally found in Ref.\ \cite{Bardeen1972}
\begin{align}
    E &= {\pm} \frac{r_c^{3/2} - 2r_c^{1/2} \pm a}
    {r_c^{3/4}\sqrt{r_c^{3/2} - 3r_c^{1/2} \pm 2a}}
    \label{eq:Ecirc}\\
    L &= \pm \frac{r_c^2 \mp 2ar_c^{1/2} + a^2}
    {r_c^{3/4}\sqrt{r_c^{3/2} - 3r_c^{1/2} \pm 2a}}\quad ,
    \label{eq:Lcirc}
\end{align}
where the $\pm$ refers to prograde and retrograde orbits respectively.

\begin{figure}
  \centering
  \includegraphics[width=0.4\textwidth]{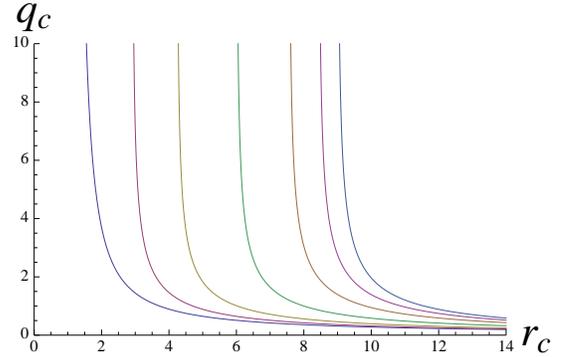}
\hfill
\caption{$q_c$ for circular orbits for spin values from left to right
$a=0.998,0.8,0.5,0,-0.5,-0.8,-0.998$. Negative spin values are
  equivalent to retrograde orbits. The value of $q_c$ diverges at the innermost stable circular orbit
  (ISCO) for each value of the spin as expected \cite{levin2008}.
\label{qcfig}}  \end{figure}

Alternatively, and perhaps 
more easily, we can write the roots as functions of $(a,E,L)$ by expanding,
\begin{align}
&R(r)/r=0\nonumber \\
&=(E^2-1)(r-r_0)(r-r_c)^2\nonumber \\
&=(E^2-1)\left (r^3-r^2(r_0+2r_c)+r(r_c^2+2r_0r_c)-r_0r_c^2\right )
\nonumber 
\end{align}
to find the $r\ne 0$ roots when $r_a=r_p$ are degenerate at the value $r_c$.
Equating coefficients with those of Eq.\ (\ref{coeff}) gives
\begin{align}
A &=(E^2-1) \nonumber \\
-A(r_0+2r_c)&=2 \nonumber \\
A(r_c^2+2r_0r_c) &= Aa^2-L^2 \nonumber \\
-Ar_0r_c^2 & = 2 (L-aE)^2 \quad ,
\end{align}
and
\begin{align}
r_c &=-\frac{2}{3(E^2-1)}\left [1+
  \sqrt{1-\frac{3(E^2-1)}{4}\left ((E^2-1)a^2-L^2\right )}\right ]
\nonumber \\
r_0 &=-\frac{2}{(E^2-1)}-2r_c
\quad .
\end{align}
When solving the quadratic, the sign consistent with $(r_c-r_0)>0$ is
chosen.
Note that for all bound orbits $E^2-1<0$.

The value of $q_c$ for different spins is shown in
Fig.\ \ref{qcfig}. In the strong-field, the ground state has a $q_c$
well above zero, ruling out Mercury-type precessions in favor of the
more extreme
zoom-whirl precessions.

\begin{comment}
A useful relation will be 
\begin{equation}
r_c-r_0=-\frac{2}{(E^2-1)}  \sqrt{1-\frac{3}{4}\left ((E^2-1)a^2-L^2\right )}
\end{equation}
for stable circular orbits.
\end{comment}

\section{Gravitational Radiation}

In addition to being fundamental,
the black hole atoms have the added benefit of being real.
Black hole pairs formed by tidal capture in dense regions, such as
globular clusters \cite{wen2003} or the galactic nucleus
\cite{Kocsis:2006hq,O'Leary:2008xt}, will likely become bound on
eccentric orbits. Light black holes that fall onto supermassive black
holes in the galactic nucleus are also free to be eccentric.
Supermassive black holes with stellar mass black hole companions
are candidates for direct detection through gravitational
radiation by LISA (Laser Interferometer Space Antenna). 
As entries in the periodic tables, we can identify
the $q$ of these black hole inspirals as a function of $(a,E,L)$ given the
results of this paper.
This allows us to predict the relativistic zoom-whirl behavior from a simple formula.

In our
periodic tables and energy level diagrams we have neglected the
dissipative effects of 
gravitational radiation. As energy and angular momentum are radiated
away in the form of gravitational waves, there will be transitions
from the orbits of one energy level diagram at a given $L$ to those of
another
energy level diagram at lower $L$. 

\begin{figure}
  \centering
  \includegraphics[width=0.4\textwidth]{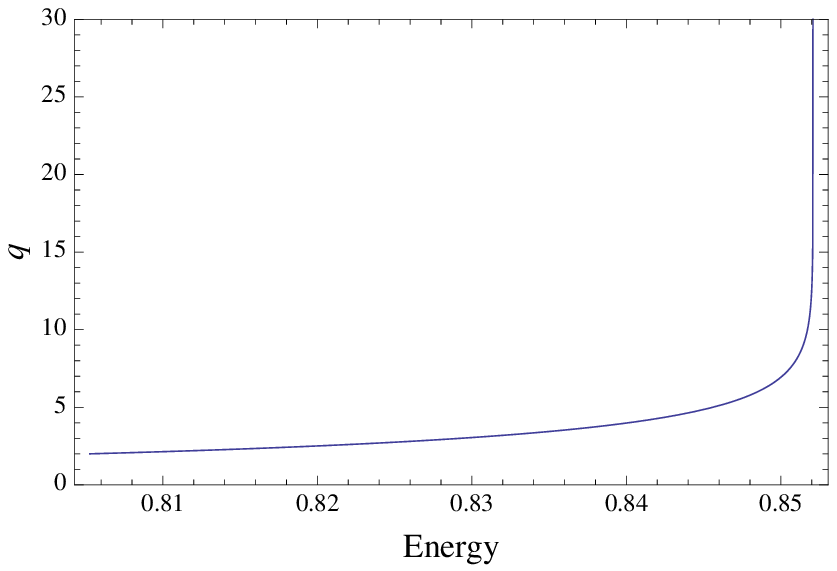}
  \includegraphics[width=0.4\textwidth]{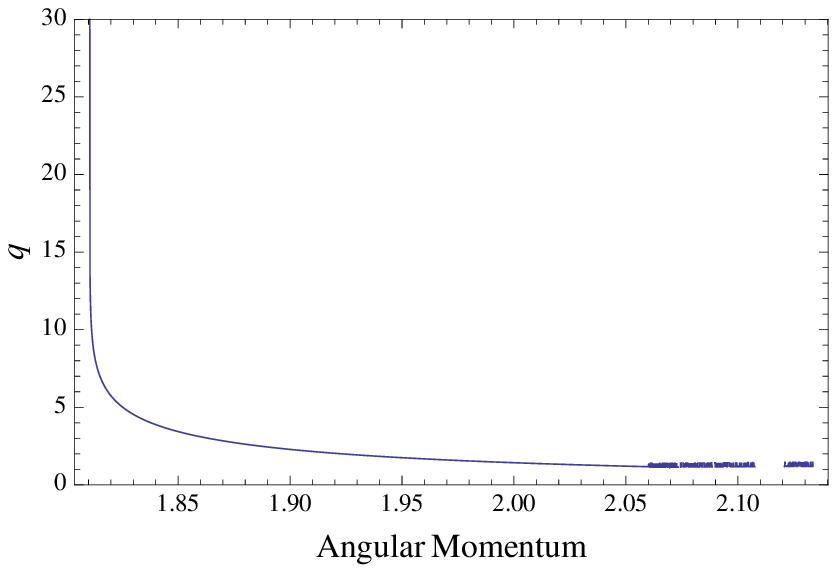}
\hfill
\caption{$a=0.995$ Top: $q$ versus $E$ for $L$ half way between the
  ibco and the isco, $L=1.81058$. Bottom: $q$ versus $L$ for $E=0.852079$.
\label{qEL}}  \end{figure}

The rate of change of $q$ can be expressed in terms of the change of
energy and angular momentum as
\begin{equation}
\frac{d q}{dt} = \frac{\partial q}{\partial E}\frac{dE}{dt}
+\frac{\partial q}{\partial L}\frac{dL}{dt} \quad .
\end{equation}
This can be evaluated by explicitly taking the derivatives of
$(\ref{qfinal})$, being sure to treat $r_0,r_p,r_a$ as functions of
$(E,L)$ through Eqs.\ (\ref{roots}). Accurate calculations of $dE/dt$
and $dL/dt$ for zoom-whirl orbits require detailed numerical computations
\cite{drasco2004,drasco2005,drasco2006}. In conjuction with such computations, our closed
form expression for $q$ would enable us to find possible resonances
during inspiral \cite{Hinderer:2008dm}, which occur during episodes of
$dq/dt\approx 0$.
As seen in Fig.\ \ref{qEL}, $\partial q/\partial E >0$ and $\partial
q/\partial L<0$. Consequently,
it is possible that the condition $\Delta q\approx 0$ will be satisfiable:
\begin{equation}
\frac{\partial q}{\partial E}\Delta E\approx - \frac{\partial
  q}{\partial L}\Delta L \quad ,
\end{equation}
during inspiral and an orbit will noticeably hang near a given $q$ for
a time. While a detailed study will require a full numerical treatment
of adiabatic inspiral, we mention that in the simulations of
Ref.\ \cite{drasco}, it can be seen by eye that the orbit will
occasionally hang on a very nearly closed orbit as it sweeps through
to merger.
 
In full numerical relativity we have seen $q$ change rapidly for
comparable mass black hole pairs \cite{Healy:2009zm}. By contrast,
an extreme-mass-ratio system can spend 
hundreds to thousands of orbits in the strong-field regime where
periodic tables are most illuminating. The small black hole will
cascade through a sequence of orbits, each of which looks like a
precession of a low-leaf clover.
As it inspirals into the black hole nucleus, it should leave an
imprint of its transit through the periodic 
table in the gravitational radiation.

%\vfill\eject
\bigskip
\noindent{*Acknowledgements*}

I thank Glenna Clifton, Becky Grossman, Jamie
Rollins and Pedro Ferreira for valuable conversations. 
I am  especially grateful to Gabe Perez-Giz
for his
important
contributions to this work.
JL gratefully
acknowledges financial support from an NSF grant
AST-0908365.

\bibliographystyle{aip}
\bibliography{gr}

\begin{thebibliography}{1}

\bibitem{chandrasekhar1983}
S.~Chandrasekhar,
\newblock {\em The Mathematical Theory of Black Holes},
\newblock Oxford: Claredon Press, 1983.

\bibitem{levin2008}
J.~Levin and G.~Perez-Giz,
\newblock Phys. Rev. D {\bf 77}, 103005 (2008).

\bibitem{2002PhRvD..66d4002G}
K.~{Glampedakis} and D.~{Kennefick},
\newblock \prd {\bf 66}, 044002 (2002).

\bibitem{carter1968}
B.~Carter,
\newblock Phys. Rev. {\bf 174}, 1559 (1968).

\bibitem{drasco2004}
S.~{Drasco} and S.~A. {Hughes},
\newblock Phys.\ Rev.\ D {\bf 69}, 044015 (2004).

\bibitem{drasco2005}
S.~Drasco, E.~Flanagan, and S.~A. Hughes,
\newblock Class.\ Quant.\ Grav. {\bf 22}, 801 (2005).

\bibitem{drasco2006}
S.~Drasco and S.~Hughes,
\newblock Phys. Rev. D {\bf 73}, 024027 (2006).

\bibitem{Hinderer:2008dm}
T.~Hinderer and E.~E. Flanagan,
\newblock (2008).

\bibitem{drasco}
S.~Drasco,
\newblock http://www.tapir.caltech.edu/~sdrasco/animations, 2009.

\end{thebibliography}

\end{document}